\documentclass{Rinton-P9x6}

\newcommand{\newc}{\newcommand} 
\newc{\beq}{\begin{equation}} 
\newc{\eeq}{\end{equation}} 
\newc{\barr}{\begin{eqnarray}} 
\newc{\earr}{\end{eqnarray}} 
\begin{document}

\title{SUSY Dark Matter Detection, The Directional Rate and the Modulation 
effect}

\author{  J. D. Vergados} 
\address{ Theoretical Physics Division, University of Ioannina, \\
E-mail: vergados@cc.uoi.gr}
\author{M. E. G\'omez}

\address{ Centro de F\'{\i}sica das 
Interac\c{c}\~{o}es Fundamentais (CFIF),  
Departamento de F\'{\i}sica, \\ Instituto Superior T\'{e}cnico, 
Av. Rovisco Pais, 1049-001 Lisboa, Portugal.\\ 
E-mail:mgomez@cfif.ist.utl.pt}

\maketitle

\abstracts{
 The detection of the theoretically expected dark matter
is central to particle physics and cosmology. Current fashionable supersymmetric
models provide a natural dark matter candidate which is the lightest
supersymmetric particle (LSP). Such models combined with fairly well 
understood physics like
the quark substructure of the nucleon and the nuclear structure (form factor 
and/or spin response function), permit the evaluation of
the event rate for LSP-nucleus elastic scattering. The thus obtained event rates
are, however, very low or even undetectable.
 So it is imperative to exploit the modulation effect, i.e. the dependence of
the event rate on  the earth's annual motion.  Also it is useful to consider
the directional rate, i.e its dependence on the direction of the recoiling
nucleus. In this paper we study
such a modulation effect both in non directional and directional experiments.
We calculate both the differential and the total rates using both isothermal, 
symmetric as well
as only axially asymmetric, and non isothermal, due to caustic rings, velocity
distributions. We find that in the symmetric case 
the modulation amplitude is small. The same is true for the case of caustic
 rings.  The inclusion of asymmetry, with a realistic enhanced velocity 
dispersion  in the galactocentric direction, yields an enhanced
modulation effect, especially in directional experiments.}
\section{Introduction}
In recent years the consideration of exotic dark matter has become necessary
in order to close the Universe \cite{Jungm}. Furthermore in
in order to understand the large scale structure of the universe 
it has become necessary to consider matter
made up of particles which were 
non-relativistic at the time of freeze out. This is  the cold dark 
matter component (CDM). The COBE data ~\cite{COBE} suggest that CDM
is at least $60\%$ ~\cite {GAW}. On the other hand during the last few years
evidence has appeared from two different teams,
the High-z Supernova Search Team \cite {HSST} and the
Supernova Cosmology Project  ~\cite {SPF} $^,$~\cite {SCP} 
 which suggests that the Universe may be dominated by 
the  cosmological constant $\Lambda$.
As a matter of fact recent data the situation can be adequately
described by  a baryonic component $\Omega_B=0.1$ along with the exotic 
components $\Omega _{CDM}= 0.3$ and $\Omega _{\Lambda}= 0.6$.
In another analysis Turner \cite {Turner} gives 
$\Omega_{m}= \Omega _{CDM}+ \Omega _B=0.4$.
Since the non exotic component cannot exceed $40\%$ of the CDM 
~\cite{Jungm},~\cite {Benne}, there is room for the exotic WIMP's 
(Weakly  Interacting Massive Particles).
  In fact the DAMA experiment ~\cite {BERNA2} 
has claimed the observation of one signal in direct detection of a WIMP, which
with better statistics has subsequently been interpreted as a modulation signal
~\cite{BERNA1}.

The above developments are in line with particle physics considerations. Thus,
in the currently favoured supersymmetric (SUSY)
extensions of the standard model, the most natural WIMP candidate is the LSP,
i.e. the lightest supersymmetric particle. In the most favoured scenarios the
LSP can be simply described as a Majorana fermion, a linear 
combination of the neutral components of the gauginos and Higgsinos
\cite{Jungm,ref1,Gomez,ref2}. 

 Since this particle is expected to be very massive, $m_{\chi} \geq 30 GeV$, and
extremely non relativistic with average kinetic energy $T \leq 100 KeV$,
it can be directly detected ~\cite{JDV96,KVprd} mainly via the recoiling
of a nucleus (A,Z) in the elastic scattering process:
\begin{equation}
\chi \, +\, (A,Z) \, \to \, \chi \,  + \, (A,Z)^* 
\end{equation}
($\chi$ denotes the LSP). In order to compute the event rate one needs
the following ingredients:

1) An effective Lagrangian at the elementary particle 
(quark) level obtained in the framework of supersymmetry as described 
, e.g., in Refs.~\cite{Jungm,ref2}.

2) A procedure in going from the quark to the nucleon level, i.e. a quark 
model for the nucleon. The results depend crucially on the content of the
nucleon in quarks other than u and d. This is particularly true for the scalar
couplings as well as the isoscalar axial coupling ~\cite{Dree}$^-$\cite{Chen}.

3) Compute the relevant nuclear matrix elements \cite{KVdubna,DIVA00}
using as reliable as possible many body nuclear wave functions.
By putting as accurate nuclear physics input as possible, 
one will be able to constrain the SUSY parameters as much as possible.
The situation is a bit simpler in the case of the scalar coupling, in which
case one only needs the nuclear form factor.

Since the obtained rates are very low, one would like to be able to exploit the
modulation of the event rates due to the earth's
revolution around the sun \cite{Verg98,Verg99}$^-$\cite{Verg01}. To this end one
adopts a folding procedure
assuming some distribution~\cite{Jungm,Verg99,Verg01} of velocities for the LSP.
One also would like to know the directional rates, by observing the 
nucleus in a certain direction, which correlate with the motion of the
sun around the center of the galaxy and the motion of the Earth 
\cite {ref1,UKDMC}.

 The calculation of this cross section  has become pretty standard.
 One starts with   
representative input in the restricted SUSY parameter space as described in
the literature~\cite{Gomez,ref2}. 
We will adopt a phenomenogical procedure taking  universal soft 
SUSY breaking terms at $M_{GUT}$, i.e., a 
common mass for all scalar fields $m_0$, a common gaugino mass 
$M_{1/2}$ and a common trilinear scalar coupling $A_0$, which 
we put equal to zero (we will discuss later the influence of 
non-zero $A_0$'s). Our effective theory below $M_{GUT}$ then 
depends on the parameters \cite{Gomez}:
\[
m_0,\ M_{1/2},\ \mu_0,\ \alpha_G,\ M_{GUT},\ h_{t},\ ,\ h_{b},\ ,\ h_{\tau},\ 
\tan\beta~,  
\]
where $\alpha_G=g_G^2/4\pi$ ($g_G$ being the GUT gauge coupling 
constant) and $h_t, h_b, h_\tau $ are respectively the top, bottom and 
tau Yukawa coupling constants at $M_{GUT}$. The values of $\alpha_G$ and 
$M_{GUT}$ are obtained as described in Ref.\cite{Gomez}.
For a specified value of $\tan\beta$ at $M_S$, we determine $h_{t}$ at 
$M_{GUT}$ by fixing the top quark mass at the center of its 
experimental range, $m_t(m_t)= 166 \rm{GeV}$. The value
of  $h_{\tau}$ at $M_{GUT}$ is  fixed by using the running tau lepton
mass at $m_Z$, $m_\tau(m_Z)= 1.746 \rm{GeV}$. 
The value of $h_{b}$ at $M_{GUT}$ used is such that:
\[
m_b(m_Z)_{SM}^{\overline{DR}}=2.90\pm 0.14~{\rm GeV}.
\]
after including the SUSY threshold correction. 
The SUSY parameter space is subject to the
 following constraints:\\
1.) The LSP relic abundance will satisfy the cosmological constrain:
\begin{equation}
0.09 \le \Omega_{LSP} h^2 \le 0.22
\label{eq:in2}
\end{equation}
{\bf 2.) The Higgs bound obtained from recent CDF \cite {VALLS} and LEP2 
\cite {DORMAN}, i.e. $m_h~>~113~GeV$.\\}
3.) We will limit ourselves to LSP-nucleon cross sections for the scalar
coupling, which gives detectable rates
\begin{equation}
4\times 10^{-7}~pb~ \le \sigma^{nucleon}_{scalar} 
\le 2 \times 10^{-5}~pb~
\label{eq:in3}
\end{equation}
 We should remember that the event rate does not depend only
 on the nucleon cross section, but on other parameters also, mainly
 on the LSP mass and the nucleus used in target. 
The condition on the nucleon cross section imposes severe constraints on the
acceptable parameter space. In particular in our model it restricts 
$tan \beta$ to values $tan \beta \simeq 50$. We will not elaborate further
on this point, since it has already appeared \cite{gtalk}. 
\bigskip
\section{Expressions for Extracting of the Nucleon Cross Section from 
the Data.} 
\bigskip

 The effective Lagrangian describing the LSP-nucleus cross section can
be cast in the form \cite {JDV96}
 \beq
{\it L}_{eff} = - \frac {G_F}{\sqrt 2} \{({\bar \chi}_1 \gamma^{\lambda}
\gamma_5 \chi_1) J_{\lambda} + ({\bar \chi}_1 
 \chi_1) J\}
 \label{eq:eg 41}
\eeq
where
\beq
  J_{\lambda} =  {\bar N} \gamma_{\lambda} (f^0_V +f^1_V \tau_3
+ f^0_A\gamma_5 + f^1_A\gamma_5 \tau_3)N~~,~~
J = {\bar N} (f^0_s +f^1_s \tau_3) N
 \label{eq:eg.42}
\eeq

We have neglected the uninteresting pseudoscalar and tensor
currents. Note that, due to the Majorana nature of the LSP, 
${\bar \chi_1} \gamma^{\lambda} \chi_1 =0$ (identically).

 With the above ingredients the differential cross section can be cast in the 
form \cite{ref1,Verg98,Verg99}
\begin{equation}
d\sigma (u,\upsilon)= \frac{du}{2 (\mu _r b\upsilon )^2} [(\bar{\Sigma} _{S} 
                   +\bar{\Sigma} _{V}~ \frac{\upsilon^2}{c^2})~F^2(u)
                       +\bar{\Sigma} _{spin} F_{11}(u)]
\label{2.9}
\end{equation}
\begin{equation}
\bar{\Sigma} _{S} = \sigma_0 (\frac{\mu_r(A)}{\mu _r(N)})^2  \,
 \{ A^2 \, [ (f^0_S - f^1_S \frac{A-2 Z}{A})^2 \, ] \simeq \sigma^S_{p,\chi^0}
        A^2 (\frac{\mu_r(A)}{\mu _r(N)})^2 
\label{2.10}
\end{equation}
\begin{equation}
\bar{\Sigma} _{spin}  =  \sigma^{spin}_{p,\chi^0}~\zeta_{spin}
\label{2.10a}
\end{equation}
\begin{equation}
\zeta_{spin}= \frac{(\mu_r(A)/\mu _r(N))^2}{3(1+\frac{f^0_A}{f^1_A})^2}
[(\frac{f^0_A}{f^1_A} \Omega_0(0))^2 \frac{F_{00}(u)}{F_{11}(u)}
  +  2\frac{f^0_A}{ f^1_A} \Omega_0(0) \Omega_1(0)
\frac{F_{01}(u)}{F_{11}(u)}+  \Omega_1(0))^2  \, ] 
\label{2.10b}
\end{equation}
\begin{equation}
\bar{\Sigma} _{V}  =  \sigma^V_{p,\chi^0}~\zeta_V 
\label{2.10c}
\end{equation}
\begin{equation}
\zeta_V =  \frac{(\mu_r(A)/\mu _r(N))^2}{(1+\frac{f^1_V}{f^0_V})^2} A^2 \, 
(1-\frac{f^1_V}{f^0_V}~\frac{A-2 Z}{A})^2 [ (\frac{\upsilon_0} {c})^2  
[ 1  -\frac{1}{(2 \mu _r b)^2} \frac{2\eta +1}{(1+\eta)^2} 
\frac{\langle~2u~ \rangle}{\langle~\upsilon ^2~\rangle}] 
\label{2.10d}
\end{equation}
\\
$\sigma^i_{p,\chi^0}=$ proton cross-section,$i=S,spin,V$ given by:\\
$\sigma^S_{p,\chi^0}= \sigma_0 ~(f^0_S)^2~(\frac{\mu _r(N)}{m_N})^2$ 
 (scalar) , 
(the isovector scalar is negligible, i.e. $\sigma_p^S=\sigma_n^S)$\\
$\sigma^{spin}_{p,\chi^0}=\sigma_0~~3~(f^0_A+f^1_A)^2~(\frac{\mu _r(N)}{m_N})^2$ 
  (spin) ,
$\sigma^{V}_{p,\chi^0}= \sigma_0~(f^0_V+f^1_V)^2~(\frac{\mu _r(N)}{m_N})^2$ 
(vector)   \\
where $m_N$ is the nucleon mass,
 $\eta = m_x/m_N A$, and
 $\mu_r(A)$ is the LSP-nucleus reduced mass,  
 $\mu_r(N)$ is the LSP-nucleon reduced mass and  
\begin{equation}
\sigma_0 = \frac{1}{2\pi} (G_F m_N)^2 \simeq 0.77 \times 10^{-38}cm^2 
\label{2.7} 
\end{equation}
\begin{equation}
u = q^2b^2/2~~,~~
Q=Q_{0}u~~, \qquad Q_{0} = \frac{1}{A m_{N} b^2} 
\label{2.15} 
\end{equation}
where
b is (the harmonic oscillator) size parameter, 
q is the momentum transfer to the nucleus, and
Q is the energy transfer to the nucleus.\\
In the above expressions $F(u)$ is the nuclear form factor and
\begin{equation}
F_{\rho \rho^{\prime}}(u) =  \sum_{\lambda,\kappa}
\frac{\Omega^{(\lambda,\kappa)}_\rho( u)}{\Omega_\rho (0)} \,
\frac{\Omega^{(\lambda,\kappa)}_{\rho^{\prime}}( u)}
{\Omega_{\rho^{\prime}}(0)} 
, \qquad \rho, \rho^{\prime} = 0,1
\label{2.11} 
\end{equation}
are the spin form factors \cite{KVprd} ($\rho , \rho^{'}$ are isospin indices)
Both form factors are normalized to one at $u=0$.
$\Omega_0$ ($\Omega_1$) are the static isoscalar (isovector) spin 
matrix elements.

 The non-directional event rate is given by:
\begin{equation}
R=R_{non-dir} =\frac{dN}{dt} =\frac{\rho (0)}{m_{\chi}} \frac{m}{A m_N} 
\sigma (u,\upsilon) | {\boldmath \upsilon}|
\label{2.17} 
\end{equation}
 Where
 $\rho (0) = 0.3 GeV/cm^3$ is the LSP density in our vicinity and 
 m is the detector mass 
The differential non-directional  rate can be written as
\begin{equation}
dR=dR_{non-dir} = \frac{\rho (0)}{m_{\chi}} \frac{m}{A m_N} 
d\sigma (u,\upsilon) | {\boldmath \upsilon}|
\label{2.18}  
\end{equation}
where $d\sigma(u,\upsilon )$ was given above.

 The directional differential rate \cite{ref1},\cite{Verg01} in the
direction $\hat{e}$ is given by :
\begin{equation}
dR_{dir} = \frac{\rho (0)}{m_{\chi}} \frac{m}{A m_N} 
{\boldmath \upsilon}.\hat{e} H({\boldmath \upsilon}.\hat{e})
 ~\frac{1}{2 \pi}~  
d\sigma (u,\upsilon)
\label{2.20}  
\end{equation}
where H the Heaviside step function. The factor of $1/2 \pi$ is 
introduced, since  the differential cross section of the last equation
is the same with that entering the non-directional rate, i.e. after
an integration
over the azimuthal angle around the nuclear momentum has been performed.
In other words, crudely speaking, $1/(2 \pi)$ is the suppression factor we
 expect in the directional rate compared to the usual one. The precise 
suppression factor depends, of course, on the direction of observation.
In spite of their very interesting experimental signatures, we will
not be concerned here with directional rates.
The mean value of the non-directional event rate of Eq. (\ref {2.18}), 
is obtained by convoluting the above expressions with the LSP velocity
distribution $f({\bf \upsilon}, {\boldmath \upsilon}_E)$ 
with respect to the Earth, i.e. is given by:
\beq
\Big<\frac{dR}{du}\Big> =\frac{\rho (0)}{m_{\chi}} 
\frac{m}{A m_N}  
\int f({\bf \upsilon}, {\boldmath \upsilon}_E) 
          | {\boldmath \upsilon}|
                       \frac{d\sigma (u,\upsilon )}{du} d^3 {\boldmath \upsilon} 
\label{3.10} 
\eeq
 The above expression can be more conveniently written as
\beq
\Big<\frac{dR}{du}\Big> =\frac{\rho (0)}{m_{\chi}} \frac{m}{Am_N} \sqrt{\langle
\upsilon^2\rangle } {\langle \frac{d\Sigma}{du}\rangle } 
\label{3.11}  
\eeq
where
\beq
\langle \frac{d\Sigma}{du}\rangle =\int
           \frac{   |{\boldmath \upsilon}|}
{\sqrt{ \langle \upsilon^2 \rangle}} f({\boldmath \upsilon}, 
         {\boldmath \upsilon}_E)
                       \frac{d\sigma (u,\upsilon )}{du} d^3 {\boldmath \upsilon}
\label{3.12}  
\eeq

 After performing the needed integrations over the velocity distribution,
to first order in the Earth's velocity, and over the energy transfer u  the
 last expression takes the form
\beq
R =  \bar{R}~t~
          [1 + h(a,Q_{min})cos{\alpha})] 
\label{3.55a}  
\eeq
where $\alpha$ is the phase of the Earth ($\alpha=0$ around June 2nd)
and  $Q_{min}$ is the energy transfer cutoff imposed by the detector.
In the above expressions $\bar{R}$ is the rate obtained in the conventional 
approach \cite {JDV96} by neglecting the folding with the LSP velocity and the
momentum transfer dependence of the differential cross section, i.e. by
\beq
\bar{R} =\frac{\rho (0)}{m_{\chi}} \frac{m}{Am_N} \sqrt{\langle
v^2\rangle } [\bar{\Sigma}_{S}+ \bar{\Sigma} _{spin} + 
\frac{\langle \upsilon ^2 \rangle}{c^2} \bar{\Sigma} _{V}]
\label{3.39b}  
\eeq
where $\bar{\Sigma} _{i}, i=S,V,spin$ have been defined above, see Eqs
 (\ref {2.10}) - (\ref {2.10c}). It contains all the parameters of the
SUSY models. 
 The modulation is described by the parameter $h$ . Once
the rate is known and the parameters $t$ and $h$, which depend only
on the LSP mass, the nuclear form factor and the velocity distribution 
the nucleon cross section can be extracted and compared to experiment.

The total  directional event rates  can be obtained in a similar fashion by
by integrating Eq. (\ref {2.20}) 
with respect to the velocity as well as the energy transfer u.  We find

\barr
R_{dir}& = &  \bar{R} [(t^0/4 \pi) \, 
            |(1 + h_1(a,Q_{min})cos{\alpha}) {\bf e}~_z.{\bf e}
\nonumber\\  &-& h_2(a,Q_{min})\, 
cos{\alpha} {\bf e}~_y.{\bf e}
                      + h_3(a,Q_{min})\, 
sin{\alpha} {\bf e}~_x.{\bf e}|
\label{4.55}  
\earr
We remind that the z-axis is in the direction of the sun's motion, the y-axis
 is perpendicular to the plane of the galaxy and the x-axis is in the 
galactocentric direction.
 The effect of folding
with LSP velocity on the total rate is taken into account via the quantity
$t^0$, which depends on the LSP mass. All other SUSY parameters have been
 absorbed in $\bar{R}$.
 We see that the modulation of the directional total event rate can
be described in terms of three parameters $h_l$, l=1,2,3. 
 In the special case of $\lambda=0$ we essentially have  one 
parameter, namely $h_1$, since then we have $h_2=0.117$ and $h_3=0.135$.
Given the functions $h_l(a,Q_{min})$ one can plot the the expression in
Eq. (\ref {4.55}) as a function of the phase of the earth $\alpha$. 
\section{The Scalar Contribution- The Role of the Heavy Quarks}
\bigskip

The coherent scattering can be mediated via the the neutral
intermediate Higgs particles (h and H), which survive as physical 
particles. It can also be mediated via s-quarks, via the mixing
of the isodoublet and isosinlet s-quarks of the same charge. In our
model we find that the Higgs contribution becomes dominant and, as a
matter of fact the heavy Higgs H is more important (the Higgs particle
$A$ couples in a pseudoscalar way, which does not lead to coherence).
It is well known that all quark flavors contribute
\cite{Dree}, since the relevant couplings are proportional to
the quark masses.
One encounters in the nucleon not only the usual 
sea quarks ($u {\bar u}, d {\bar d}$ and $s {\bar s}$) but the 
heavier quarks $c,b,t$ which couple to the nucleon via two gluon 
exchange, see e.g. Drees {\it et al}  ~\cite{Dree00} and references
therein.
 
As a result  one obtains an effective scalar Higgs-nucleon
coupling by using  effective quark masses as follows
\begin{center}
$m_u \ra f_u m_N, \ \ m_d \ra f_d m_N. \ \ \  m_s \ra f_s m_N$   
\end{center}
\begin{center}
$m_Q \ra f_Q m_N, \ \ (heavy\ \  quarks \ \ c,b,t)$   
\end{center}
where $m_N$ is the nucleon mass. The isovector contribution is now
negligible. The parameters $f_q,~q=u,d,s$ can be obtained by chiral
symmetry breaking 
terms in relation to phase shift and dispersion analysis.
Following Cheng and Cheng ~\cite{Chen} we obtain:
\begin{center}
$ f_u = 0.021, \quad f_d = 0.037, \quad  f_s = 0.140$ 
\quad  \quad  (model B)   
\end{center}
\begin{center}
$ f_u = 0.023, \quad f_d = 0.034, \quad  f_s = 0.400$ 
\quad  \quad  (model C)   
\end{center}
 We see that in both models the s-quark is dominant.
Then to leading order via quark loops and gluon exchange with the
nucleon one finds:
\begin{center}
\quad $f_Q= 2/27(1-\quad \sum_q f_q)$   
\end{center}
This yields:
\begin{center}
\quad $ f_Q = 0.060$    (model B),   
\quad $ f_Q = 0.040$    (model C)   
\end{center}
 There is a correction to the above parameters coming from loops
involving s-quarks \cite {Dree00} and due to QCD effects. 
 Thus for large $tan \beta$ we find \cite {ref1}:
\begin{center}
\quad $f_{c}=0.060 \times 1.068=0.064,
       f_{t}=0.060 \times 2.048=0.123,
       f_{b}=0.060 \times 1.174=0.070$  \quad (model B)
\end{center}
\begin{center}
\quad $f_{c}=0.040 \times 1.068=0.043,
       f_{t}=0.040 \times 2.048=0.082,
       f_{b}=0.040 \times 1.174=0.047$  \quad (model B)
\end{center}
For a more detailed discussion we refer the reader to 
Refs~\cite{Dree,Dree00}.
\section{Results and Discussion}
\bigskip

\begin{figure}
\epsfig{figure=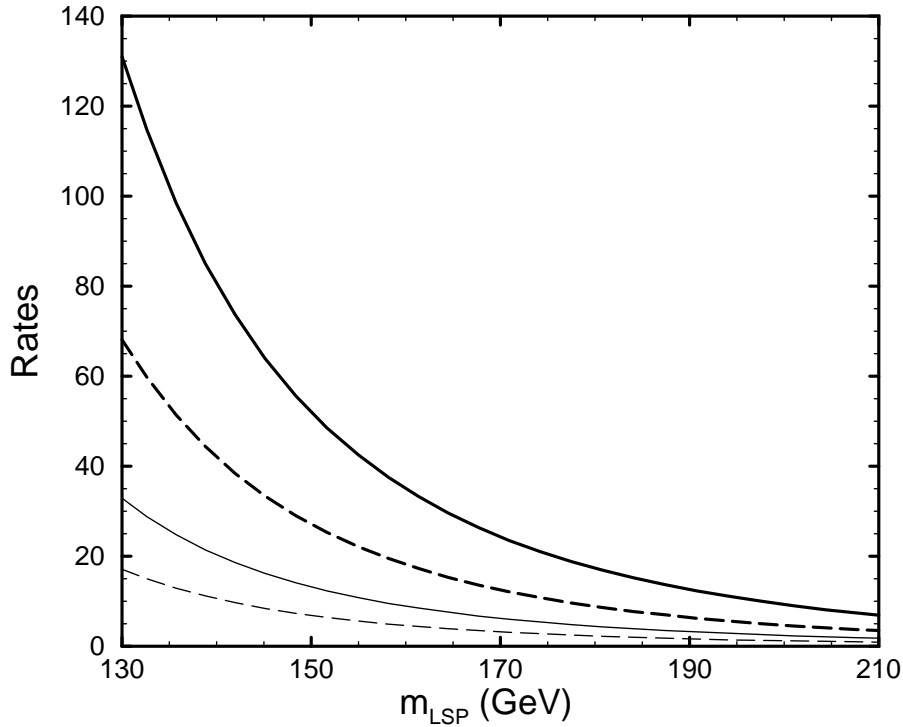,height=3.90in,angle=0}
\caption{The Total detection rate per $(kg-target)yr$ vs the LSP mass
in GeV for a typical solution in our parameter space in the case of 
$^{127}I$ corresponding to  model B (thick line) and Model C (fine 
line). For the definitions see text.
\label{rate}}
\end{figure}

The three basic ingredients of our calculation were 
the input SUSY parameters (see sect. 1), a quark model for the nucleon
(see sect. 3) and the velocity distribution combined with the structure of
 the nuclei involved (see sect. 2). we will focus our attention on the
coherent scattering and present results for the popular target $^{127}I$.
We have utilized two nucleon models indicated by B and C which take into
account the presence of heavy quarks in the nucleon. We also considered
energy cut offs imposed by the detector,  
 by considering two  typical cases $Q_{min}=10,~20$ KeV. The thus obtained
results for the unmodulated total non directional event rates $\bar{R}t$
 in the case
 of the symmetric isothermal model for a typical SUSY parameter choice
\cite{Gomez} are shown in Fig. \ref{rate}. 
Special attention was paid to the the directional rate and its  
modulation due to the annual motion of the earth in the case of isothermal
models. The case of non isothermal models, e.g. caustic rings, is more
complicated \cite{Verg01} and it will not be further discussed here.
 As expected, the parameter $t_0$, which contains the 
effect of the nuclear form factor and the LSP velocity dependence,
decreases as the reduced mass increases.
 
  We will concentrate in the case of isothermal models we will limit 
and restrict ourselves to the discussion
of the directional rates. In the special case of the direction of observation 
being close to the coordinate axes the rate is described in terms of $t_0$
and the three
quantities $h_i,~i=1,2,3$ (see Eq.  (\ref{4.55})). 
 As expected, the parameter $t_0$,
decreases as the reduced mass increases.
\begin{table}[t]  
\caption{The quantities $t^{0}$ and $h_1$ for $\lambda=0$ in the case of the
target $_{53}I^{127}$ for various LSP masses and $Q_{min}$ in KeV (for
definitions see text). Only the scalar contribution is considered. Note that in
this case $h_2$ and $h_3$ are constants equal to 0.117 and 0.135 respectively.}
\begin{center}
\footnotesize
\begin{tabular}{|l|c|rrrrrrr|}
\hline
& & & & & & & &     \\
&  & \multicolumn{7}{|c|}{LSP \hspace {.2cm} mass \hspace {.2cm} in GeV}  \\ 
\hline 
& & & & & & & &     \\
Quantity &  $Q_{min}$  & 10  & 30  & 50  & 80 & 100 & 125 & 250   \\
\hline 
& & & & & & & &     \\
$t^0$ &0.0&1.960&1.355&0.886&0.552&0.442&0.360&0.212 \\
$h_1$ &0.0&0.059&0.048&0.037&0.029&0.027&0.025&0.023 \\
\hline 
& & & & & & & &     \\
$ t^0$ &10.&0.000&0.365&0.383&0.280&0.233&0.194&0.119 \\
$h_1$ &10.&0.000&0.086&0.054&0.038&0.033&0.030&0.025 \\
\hline 
& & & & & & & &     \\
$t^0$ &20.&0.000&0.080&0.153&0.136&0.11&0.102&0.065 \\
$h_1$ &20.&0.000&0.123&0.073&0.048&0.041&0.036&0.028 \\
\hline
\hline
\end{tabular}
\end{center}
\end{table}
\begin{table}[t]  
\caption{The same as in the previous, but for the value of the asymmetry 
parameter $\lambda=1.0$.}
\begin{center}
\footnotesize
\begin{tabular}{|l|c|rrrrrrr|}
\hline
& & & & & & & &     \\
&  & \multicolumn{7}{|c|}{LSP \hspace {.2cm} mass \hspace {.2cm} in GeV}  \\ 
\hline 
& & & & & & & &     \\
Quantity &  $Q_{min}$  & 10  & 30  & 50  & 80 & 100 & 125 & 250   \\
\hline 
& & & & & & & &     \\
$t^0$ & 0.0 &2.429&1.825&1.290&0.837&0.678&0.554&0.330\\
$h_1$ & 0.0 &0.192&0.182&0.170&0.159&0.156&0.154&0.150 \\
$h_2$ & 0.0 &0.146&0.144&0.141&0.139&0.139&0.138&0.138\\
$h_3$ & 0.0 &0.232&0.222&0.211&0.204&0.202&0.200&0.198\\
\hline 
& & & & & & & &     \\
$t^0$ & 10. &0.000&0.354&0.502&0.410&0.349&0.295&0.184\\
$h_1$ & 10. &0.000&0.241&0.197&0.174&0.167&0.162&0.154\\
$h_2$ & 10. &0.000&0.157&0.146&0.142&0.140&0.140&0.138\\
$h_3$ & 10. &0.000&0.273&0.231&0.213&0.208&0.205&0.200\\
\hline 
& & & & & & & &     \\
$t^0$ & 20. &0.000&0.047&0.169&0.186&0.170&0.150&0.100\\
$h_1$ & 20. &0.000&0.297&0.226&0.190&0.179&0.172&0.159\\
$h_2$ & 20. &0.000&0.177&0.153&0.144&0.142&0.141&0.139\\
$h_3$ & 20. &0.000&0.349&0.256&0.224&0.216&0.211&0.203\\
\hline
\hline
\end{tabular}
\end{center}
\end{table}
These are shown in tables 1-2 for various values of $Q_{min}$ and $\lambda$. 
 $h_2$ and $h_3$ 
are constant, 0.117 and 0.135 respectively, in the symmetric case. On the other
hand $h_1,h_2$ and $h_3$ substantially increase in the presence of asymmetry. 
For the differential rate the reader is 
referred to our previous work \cite {Verg99,Verg00}.

\section{Conclusions}
\bigskip
In the present paper we have discussed the parameters, which describe
the event rates for direct detection of SUSY dark matter. It is well
known \cite{ref1}
the event rates are quite small and only in a small segment of the 
allowed parameter space they are above the present experimental goals.
 One, therefore, is looking for
characteristic signatures, which will aid the experimentalists in reducing
background. These are two: a) The non directional event rates, which are
correlated with the motion of the Earth (modulation effect) and b)
 the directional event rates,
which are correlated with both the velocity of the sun, around the center of
 the Galaxy, and the velocity of the Earth.
 We separated our discussion into
two parts. The first deals with the elementary aspects, SUSY parameters and
nucleon structure, and is given in terms of the nucleon cross-section. The
second deals with nucleon and nuclear effects. 
In the second step we also studied the dependence of the rates on the
energy cut off imposed by the detector. 

 Even though the actual velocity dependence may arise out of a combination of 
isothermal  and non isothermal contributions \cite{Verg01},
in the present paper we focused on isothermal models. Detectable rates are
possible with some choices of the input parameters.
A typical graph for the total unmodulated rate is shown Fig. \ref{rate}. 
We will concentrate more on the directional rates, which are described in
terms of the parameters $t_0,h_1,h_2$ and $h_3$.
To simplify matters these parameters are given in Tables 1-2 for directions
of observation close to the three axes $x,y,z$.  We see that the unmodulated 
rate scales by the $cos\theta_s$, with $\theta_s$ being the angle between
the direction of observation and the velocity of the sun, The reduction factor
 of the total directional
rate, along the sun's direction of motion, compared to the total non
directional rate depends, of course, on the
nuclear parameters, the reduced mass and the asymmetry parameter $\lambda$
\cite{Verg00}.
It is given by the parameter $f_{red}=t_0/(4 \pi~t_0)=\kappa/(2 \pi)$. We find
that $\kappa$ is around 0.6 for no asymmetry and around 0.7 for maximum
asymmetry ($\lambda=1.0$). In other words it is not very different from the 
naively expected $f_{red}=1/( 2 \pi)$, i.e.  $\kappa=1$. 
The modulation of the directional rate 
increases with the asymmetry parameter $\lambda$ and it depends, of
course, on
 the direction of observation. For $Q_{min}=0$ it can reach values up 
to $23\%$. Values up to $35\%$ are possible for large values of
 $Q_{min}$, but they occur at the expense of the total
number of counts.

\par
This work was supported by the European Union under the contracts 
RTN No HPRN-CT-2000-00148 and TMR 
No. ERBFMRX--CT96--0090 and $\Pi E N E \Delta~95$ of the Greek 
Secretariat for Research.

\def\ijmp#1#2#3{{ Int. Jour. Mod. Phys. }{\bf #1~}(#2)~#3}
\def\pl#1#2#3{{ Phys. Lett. }{\bf B#1~}(#2)~#3}
\def\zp#1#2#3{{ Z. Phys. }{\bf C#1~}(#2)~#3}
\def\prl#1#2#3{{ Phys. Rev. Lett. }{\bf #1~}(#2)~#3}
\def\rmp#1#2#3{{ Rev. Mod. Phys. }{\bf #1~}(#2)~#3}
\def\prep#1#2#3{{ Phys. Rep. }{\bf #1~}(#2)~#3}
\def\pr#1#2#3{{ Phys. Rev. }{\bf D#1~}(#2)~#3}
\def\np#1#2#3{{ Nucl. Phys. }{\bf B#1~}(#2)~#3}
\def\npps#1#2#3{{ Nucl. Phys. (Proc. Sup.) }{\bf B#1~}(#2)~#3}
\def\mpl#1#2#3{{ Mod. Phys. Lett. }{\bf #1~}(#2)~#3}
\def\arnps#1#2#3{{ Annu. Rev. Nucl. Part. Sci. }{\bf
#1~}(#2)~#3}
\def\sjnp#1#2#3{{ Sov. J. Nucl. Phys. }{\bf #1~}(#2)~#3}
\def\jetp#1#2#3{{ JETP Lett. }{\bf #1~}(#2)~#3}
\def\app#1#2#3{{ Acta Phys. Polon. }{\bf #1~}(#2)~#3}
\def\rnc#1#2#3{{ Riv. Nuovo Cim. }{\bf #1~}(#2)~#3}
\def\ap#1#2#3{{ Ann. Phys. }{\bf #1~}(#2)~#3}
\def\ptp#1#2#3{{ Prog. Theor. Phys. }{\bf #1~}(#2)~#3}
\def\plb#1#2#3{{ Phys. Lett. }{\bf#1B~}(#2)~#3}
\def\apjl#1#2#3{{ Astrophys. J. Lett. }{\bf #1~}(#2)~#3}
\def\n#1#2#3{{ Nature }{\bf #1~}(#2)~#3}
\def\apj#1#2#3{{ Astrophys. Journal }{\bf #1~}(#2)~#3}
\def\anj#1#2#3{{ Astron. J. }{\bf #1~}(#2)~#3}
\def\mnras#1#2#3{{ MNRAS }{\bf #1~}(#2)~#3}
\def\grg#1#2#3{{ Gen. Rel. Grav. }{\bf #1~}(#2)~#3}
\def\s#1#2#3{{ Science }{\bf #1~}(19#2)~#3}
\def\baas#1#2#3{{ Bull. Am. Astron. Soc. }{\bf #1~}(#2)~#3}
\def\ibid#1#2#3{{ ibid. }{\bf #1~}(19#2)~#3}
\def\cpc#1#2#3{{ Comput. Phys. Commun. }{\bf #1~}(#2)~#3}
\def\astp#1#2#3{{ Astropart. Phys. }{\bf #1~}(#2)~#3}
\def\epj#1#2#3{{ Eur. Phys. J. }{\bf C#1~}(#2)~#3}


\begin{thebibliography}{99}
\bibitem{Jungm}For a review see e.g.
G. Jungman {\it et al.},{\it  Phys. Rep.} {\bf 267}, 195  (1996).
\bibitem{COBE}G.F. Smoot et al., (COBE data), {\it Astrophys. J.} {\bf 396} 
(1992) L1.
\bibitem{GAW}E. Gawiser and J. Silk,{\it {Science}} {\bf 280}, 1405 (1988);
M.A.K. Gross, R.S. Somerville, J.R. Primack, J. Holtzman and A.A. Klypin,
{\it Mon. Not. R. Astron. Soc.} {\bf 301}, 81 (1998).
\bibitem{HSST}A.G. Riess {\it et al}, {\it Astron. J.}
{\bf 116} (1998), 1009.
\bibitem{SPF} R.S. Somerville, J.R. Primack and S.M. Faber, astro-ph/9806228;
{\it Mon. Not. R. Astron. Soc.} (in press).
\bibitem{SCP}Perlmutter, S. {\it et al} (1999) {\it Astrophys. J.}
{\bf 517},565; (1997) {\bf 483},565 ({\it astro-ph}/9812133).\\
S. Perlmutter, M.S. Turner and M. White, {\it Phys. Rev. Let.} {\bf 83},
670 (1999).
\bibitem{Turner} {\it Cosmological parameters},{\it astro-ph}/9904051; 
{\it Phys. Rep.} {\bf 333-334} (1990), 619.
\bibitem{Benne} D.P. Bennett {\it et al.}, (MACHO collaboration), A binary
lensing event toward the LMC: Observations and Dark Matter Implications, 
Proc. 5th Annual Maryland Conference, edited by S. Holt (1995);\\
C. Alcock {\it et al.}, (MACHO collaboration), {\it Phys. Rev. Lett.} {\bf 74}
, 2967 (1995). 
\bibitem{BERNA2} R. Bernabei et al., INFN/AE-98/34, (1998);
R. Bernabei et al., {it Phys. Lett.} {\bf B 389}, 757 (1996).
\bibitem{BERNA1} R. Bernabei et al., {\it Phys. Lett.} {\bf B 424}, 195 (1998);
{\bf B 450}, 448 (1999).
\bibitem{ref1}For more references see e.g. our previous report:\\
J.D. Vergados, Supersymmetric Dark Matter Detection-
The Directional Rate and the Mmodulation Effect, hep-ph/0010151;
\bibitem{Gomez} M.E. G\'{o}mez, J.D. Vergados, hep-ph/0012020.\\
M.E. G\'{o}mez, G. Lazarides and C. Pallis, \pr{61}{2000}{123512} and
 {\it Phys. Lett.} {\bf B 487}, 313 (2000).
\bibitem{gtalk} M.E. G\'{o}mez and  J.D. Vergados, hep-ph/0105115.
\bibitem{ref2}
A. Bottino {\it et al.}, {\it Phys. Lett} {\bf B 402}, 113 (1997).\\
R. Arnowitt and P. Nath, {\it Phys. Rev. Lett.}  {\bf 74}, 4952 (1995);
 {\it Phys. Rev.} {\bf D 54}, 2394 (1996); hep-ph/9902237;\\
V.A. Bednyakov, H.V. Klapdor-Kleingrothaus and S.G. Kovalenko,
{\it Phys. Lett.}  {\bf B 329}, 5 (1994).
\bibitem{JDV96}J.D. Vergados, {\it J. of Phys.} {\bf G 22}, 253 (1996).
\bibitem{KVprd}T.S. Kosmas and J.D. Vergados, {\it Phys. Rev.} {\bf D 55}, 
1752 (1997).
{\it Phys. Lett.}  {\bf B 329}, 5 (1994).
\bibitem{drees} M. Drees and M. M. Nojiri, 
\pr{47}{1993}{376}; 
\bibitem{Dree}M. Drees and M.M. Nojiri, {\it Phys. Rev.} {\bf D 48}, 
3843 (1993); {\it Phys. Rev.} {\bf D 47}, 4226 (1993). 
\bibitem{Dree00}A. Djouadi and M. Drees, Phys.\ Lett.\ B {\bf 484} (2000) 183;
S. Dawson, {\it Nucl. Phys.} {\bf B359},283 (1991);
M. Spira {it et al}, {\it Nucl. Phys.} {\bf B453},17 (1995).
\bibitem{Chen}T.P. Cheng, {\it Phys. Rev.} {\bf D 38} 2869 (1988);
H -Y. Cheng, Phys. Lett. {\bf B 219} 347 (1989).
\bibitem{KVdubna}J.D. Vergados and T.S. Kosmas, {\it Physics of Atomic
 nuclei}, Vol. {\bf 61}, No 7, 1066 (1998) 
(from {\it Yadernaya Fisika}, Vol. 61, No 7, 1166 (1998).
\bibitem{DIVA00}P.C. Divari, T.S. Kosmas, J.D. Vergados and L.D. Skouras,
{\it  Phys. Rev.} {\bf C 61} (2000), 044612-1.
\bibitem{Verg98}J.D. Vergados, {\it Phys. Rev.} {\bf D 58}, 103001-1 (1998).
\bibitem{Verg99}J.D. Vergados, {\it Phys. Rev. Lett} {\bf  83}, 3597
(1999).
\bibitem{Verg00}J.D. Vergados, {\it Phys. Rev.} {\bf D 62}, 023519 (2000).
\bibitem{Verg01}J.D. Vergados, {\it Phys. Rev. } {\bf D 63}, 06351 (2001).
i
\bibitem{UKDMC}K.N. Buckland, M.J. Lehner, G.E. Masek, in 
 Proc. {\it 3nd Int. Conf. on Dark Matter
in Astro- and part. Phys.} (Dark2000), Ed. H.V. Klapdor-Kleingrothaus,
Springer Verlag (2000).
\bibitem{VALLS} {\it CDF Collaboration}, FERMILAB-Conf-99/263-E CDF;\\
http://fnalpubs.fnal.gov/archive/1999/conf/Conf-99-263-E.html.
\bibitem{DORMAN} P.J. Dorman, {\it ALEPH Collaboration}, March 2000,\\
http://alephwww.cern.ch/ALPUB/seminar/lepc$_{}$mar200/lepc2000.pdf.
\end{thebibliography}
\end{document}